\documentclass[journal]{IEEEtran}
\usepackage{cite}

\ifCLASSINFOpdf
  \usepackage[pdftex]{graphicx}
\else
\fi
\usepackage{float}
\usepackage{amsfonts}
\usepackage{subfiles}

\usepackage{amsmath}
\usepackage{amssymb}
\usepackage{amsthm}
\newtheorem{theorem}{Theorem}

\newtheorem{proposition}[theorem]{Proposition}

\newtheorem{definition}{Definition}

\usepackage{algorithmic}
\usepackage{algorithm}

\usepackage{array}

\ifCLASSOPTIONcompsoc
 \usepackage[caption=false,font=normalsize,labelfont=sf,textfont=sf]{subfig}
\else
 \usepackage[caption=false,font=footnotesize]{subfig}
\fi
\usepackage{url}

\usepackage{times}
\usepackage{epsfig}

\usepackage{comment}
\usepackage{makecell}
\usepackage{soul}
\usepackage{multirow}
\usepackage{booktabs}
\usepackage{enumerate}

\usepackage[pagebackref=true,breaklinks=true,colorlinks,bookmarks=false]{hyperref}

\usepackage{authblk}

\usepackage{caption}
\usepackage{color}
\usepackage{parskip}
\usepackage{indentfirst} 
\newcommand{\zyt}[1]{{\color{black}{  }}}

\begin{document}
%
\title{Morphology Edge Attention Network and Optimal Geometric Matching Connection model for vascular segmentation}
%
%
%

\author{Yuntao Zhu
\thanks{Yuntao Zhu, Yuxuan Qiao, and Xiaoping Yang are with the Department of Mathematics, Nanjing University, Nanjing 210093, P. R. China,
 e-mail: YuntaoZhu7@smail.nju.edu.cn.}
, Yuxuan Qiao, and Xiaoping Yang
\thanks{Manuscript received April xx, 20xx; revised August xx, 20xx.}}

%
%

\markboth{Journal of \LaTeX\ Class Files,~Vol.~xx, No.~x, August~20xx}%
{Shell \MakeLowercase{\textit{et al.}}: Bare Demo of IEEEtran.cls for IEEE Journals}
%



\maketitle

\begin{abstract}
There are many unsolved problems in vascular image segmentation, including vascular structural connectivity, scarce branches and missing small vessels.
Obtaining vessels that preserve their correct topological structures is currently a crucial research issue, as it provides an overall view of one vascular system.
In order to preserve the topology and accuracy of vessel segmentation, 
we proposed a novel Morphology Edge Attention Network (MEA-Net) for the segmentation of vessel-like structures,
and an Optimal Geometric Matching Connection (OGMC) model to connect the broken vessel segments.
The MEA-Net has an edge attention module that improves the segmentation of edges and small objects by morphology operation extracting boundary voxels on multi-scale. 
The OGMC model uses the concept of curve touching from differential geometry to filter out fragmented vessel endpoints, and then employs minimal surfaces to determine the optimal connection order between blood vessels.
Finally, we calculate the geodesic to repair missing vessels under a given Riemannian metric.
Our method achieves superior or competitive results compared to state-of-the-art methods on four datasets of 3D vascular segmentation tasks, both effectively reducing vessel broken and increasing vessel branch richness, yielding blood vessels with a more precise topological structure.

\end{abstract}


%
\IEEEpeerreviewmaketitle

\section{Introduction}
\begin{figure}
    \centering
    \includegraphics[width= 0.5\linewidth]{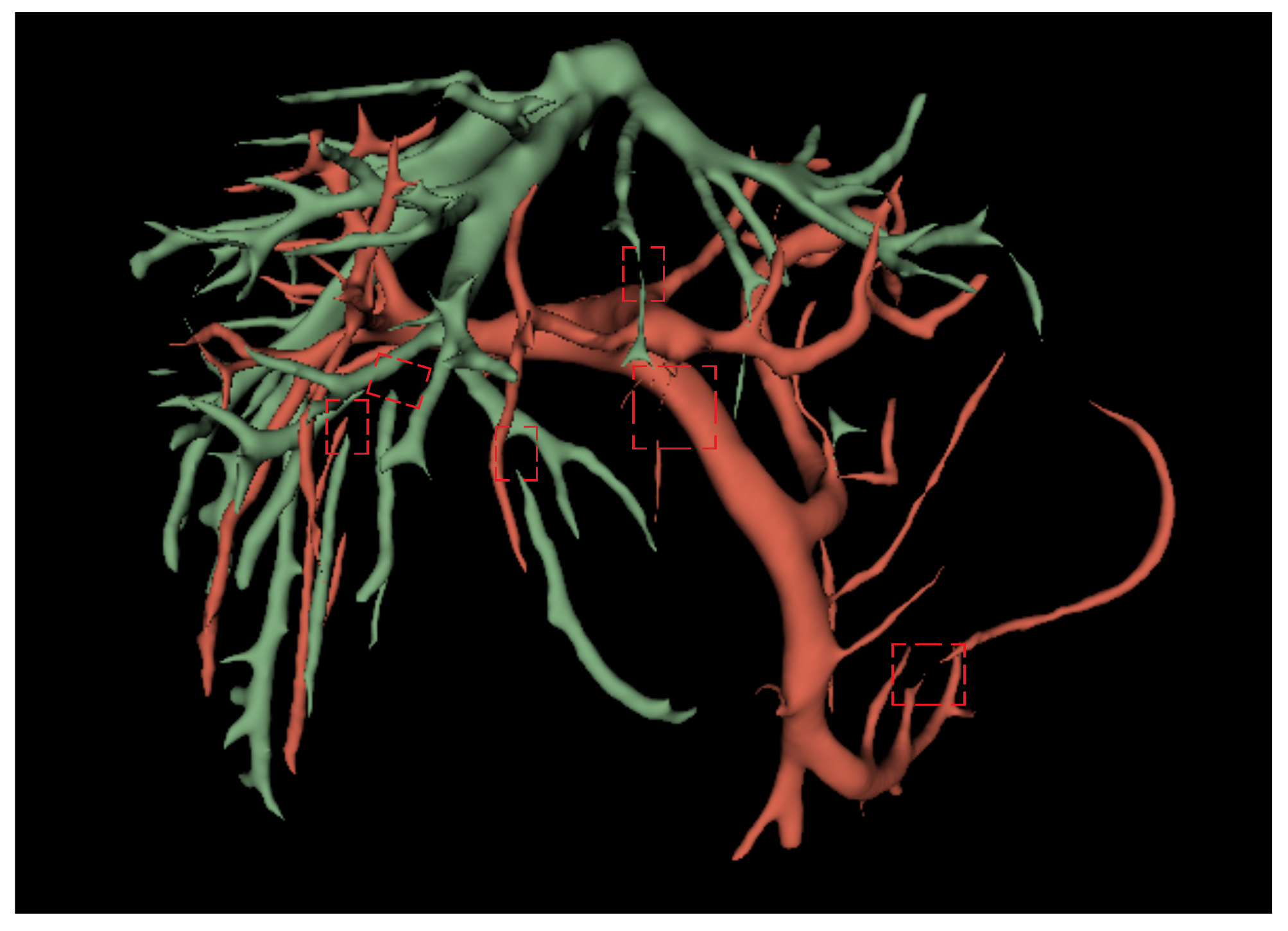}
    \caption{
    Example of liver vessel fragments.
    }
    \label{fig:data+fragment}
\end{figure}

\bstctlcite{IEEEexample:BSTcontrol} 
%
%
%
%
\IEEEPARstart{A}{ccurate} 
    vessel segmentation is crucial for the treatment of various diseases and quantitative analysis \cite{mouDenseDilatedNetwork2020, huangRobustLiverVessel2018, chanRetinalVasculatureGlaucoma2017, ningChineseGuidelinePrimary2021, EASLClinicalPractice2016}. 
However, manual annotation of vascular-like structures is a time-consuming and labour-intensive process. 
Furthermore, most existing computer-aided systems struggle to reliably extract and segment these structures due to issues such as varying levels of noises, low image contrast, and complex variations in vessel geometry and topology. These challenges often result in vessel fractures, as shown in Figure \ref{fig:data+fragment}.

Traditional vessel segmentation methods, including vessel enhancement filtering \cite{frangiMultiscaleVesselEnhancement1998,jermanEnhancementVascularStructures2016}, region growth \cite{jiangRegionGrowingVessel2013a}, active contour model \cite{yanfengshangVascularActiveContour2011}, graph cut model \cite{zhaiLungVesselSegmentation2016a}, and minimal path model \cite{benmansourFastObjectSegmentation2009, chenGlobalMinimumFinsler2017}, had unsatisfactory accuracy, especially in cases where the vessel are difficult to distinguish from the surrounding body tissues 
\cite{frangiMultiscaleVesselEnhancement1998, jermanEnhancementVascularStructures2016,jiangRegionGrowingVessel2013a,yanfengshangVascularActiveContour2011,zhaiLungVesselSegmentation2016a,benmansourFastObjectSegmentation2009,chenGlobalMinimumFinsler2017}.
With the rapid development of deep learning, many algorithms based network have been proposed for vessel detection and have achieved state-of-the-art results \cite{mookiahReviewMachineLearning2021,ciecholewskiComputationalMethodsLiver2021,guptaLearningTopologicalInteractions2022a}.
Although these deep learning-based methods show better performance, they often fail to detect complex geometric and topological variations or tiny vessels, leading to vessel fractures \cite{mookiahReviewMachineLearning2021,ciecholewskiComputationalMethodsLiver2021,guptaLearningTopologicalInteractions2022a}.

   Currently, there are two kind of methods aim to achieve segmentation results with fewer breaks and more accurate topology: end-to-end and two-stage approaches.
Among the end-to-end methods, Clough et al. \cite{cloughTopologicalLossFunction2020} and Hu et al. \cite{hu2021topologyaware} utilized algebraic topology theory to compute topological features of vessels, such as Betti numbers or persistent homotopy groups, to regularize the network. On the other hand, Shit et al. \cite{shitClDiceNovelTopologyPreserving2021}, Oner et al. \cite{onerEnforcingConnectivity3D2022} and Hu et al. \cite{huImageSegmentationHomotopy2021} deformed the vessel mask by extracting the vessel centreline or performing homotopy transformations. They then constructed loss functions that guide the network to detect the transformed ground truth.
Various attention modules \cite{wangNonlocalNeuralNetworks2018, huSqueezeandexcitationNetworks2018, vaswaniAttentionAllYou2017a} have been proposed to improve the ability of the network to detect specific vessel parts. For example, Xia et al. \cite{xia3DVessellikeStructure2022} introduced the Reverse Edge Attention Module (REAM), which uses high-level features to construct a low-level reverse map and captures edge information by looking for the difference in foreground. Furthermore, Gupta et al. \cite{guptaLearningTopologicalInteractions2022a} proposed multi-class interactions to impose edge constraints, such as vessel-wall interactions, by exploiting object edge interaction relations of inclusion or exclusion to emphasise the weight of edge pixels.
However, these methods often neglect to exploit local geometric structure features and the global relationship between vessels.

    Two-stage methods aim to repair the results of network segmentation by identifying and reconnecting the missing parts. 
    Zhang et al. \cite{zhangConfluentVesselTrees2021} transformed the entire vessel into a graph representation and used a minimum spanning tree algorithm to generate a new vessel that reconnects the broken segments.
    Mou et al. \cite{mouDenseDilatedNetwork2020} utilized a curve representing the broken peripheral vessels and expand it to replace the missing vessel segments. 
    Other methods, such as those proposed by Benmansour et al. \cite{benmansourFastObjectSegmentation2009}, Chen et al. \cite{chenVesselTreeExtraction2016}, Liu et al. \cite{liuTrajectoryGroupingCurvature2022} and He et al. \cite{heLearningHybridRepresentations2020a}, constructed potential maps. They then used the minimal path method to connect the two endpoints of a given broken vessel. 
    However, these methods had not considered geometric and pair priority in a unified framework and could not been applied to 3-dimensional situations.

In this article,  we propose a morphology edge attention network (MEA-Net) and an optimal geometry matching connection (OGMC) model for vessel segmentation.
Figure \ref{fig:overall} shows the overall architecture of the proposed method.
MEA-Net enables to enhance neural network ability to distinguish tiny vessels and vascular edges by a differentiable morphological operator to extract edges, and it contains no other trainable parameters and interpretation.
Based on the prediction mask given by MEA-Net, OGMC model is designed to use touching fit degree (TFD) to determine which are breakpoint pairs in the prediction mask, and introduce minimal surface as a solution how to optimally connect all candidate point pairs.
Finally, our method prediction both increase vessel branch richness and provide blood vessels with more precise topological structures.

Our main contributions are summarized as follows:
\begin{itemize}
\item We propose a morphology edge attention network, which contains an explained mechanism of edge attention by a differentiable morphological operator to extract edges. 
\item We propose an optimal geometric matching connection model, which is a framework that includes the local geometry structure and the global relationship between vessels. In particular, it is able to preserve the topological structure of vessel-like objects.
    \item This work is the first, to the best of our knowledge, method to introduce touching and minimal surface theory from differential geometry to vessel segmentation, these concept of geometry enable us to fully exploit geometry information in vessel. 
\item Experimental results show that the proposed method improves the 
absolute Errors for the Betti Numbers
(Betti Error), 
absolute error rate of tree length (LR Error) and The absolute error rate of the branches (BR Error)
by over 10\% compared to SOTA methods and competitive accuracy on Dice Similarity Coefficient (DSC) and Normalized Symmetric Difference (NSD). The results show that our algorithm is able to effectively repair blood vessel fractures and to obtain blood vessels with the correct topological structures.
\end{itemize}

\subsection{Vessel segmentation of deep learning}

The U-Net \cite{UNet2015} and its variants \cite{xia3DVessellikeStructure2022, wuSCSNetScaleContext2021, attentionUNet} are the popular architectures in medical image segmentation.
There is a lot of work being invested in how to encourage the U-Net-like networks to focus on the vessel topology in order to produce segmentation results with less fragments. 

Clough et al.
\cite{cloughTopologicalLossFunction2020} used a persistent homology approach to observe the birth and death time of the topological structure and the Betti numbers, which change with the threshold to ensure consistency with the reference mask. 
Hu et al.
\cite{hu2021topologyaware} used the Moose 
theory to extract topological features to guide the network.

Various methods have been proposed for extracting the vessel centerline, for the centerline is a simple and effective representation of the vessel geometry. 
Lee \cite{LEE1994462} thinned a segmented vessel to extract skeleton and preserve topology invariants. 
In addition, a differentiable centreline extraction algorithm using max-min pool was introduced in \cite{shitClDiceNovelTopologyPreserving2021}.
Zhang et al.\cite{zhangDivergencePriorVesseltree2019} proposed to obtain internal voxel point orientation estimates using direction extraction algorithms to obtain vessel centrelines. 
Tetteh et al.\cite{tettehDeepVesselNetVesselSegmentation2020} used neural networks to predict the vascular centrelines.
Based on the centerline or skeleton, we can build a mathematical model to represent the vessel, and then introduce more theory to be applied in the segmentation of tubular objects.


 



\section{Preliminaries}

In this section, we introduce the concepts of touching of curves and minimal surfaces. 
\subsection{touching}
A curve $C$ in the 3-dimensional Euclidean space $E^3$ is a continuous map, denoted as:
\begin{equation}
    C:[a,b] \rightarrow E^3.
\end{equation}
Given an orthogonal frame $\{ O; \mathbf{i, j, k}\}$,
a curve $C$ can be represented as follows:
\begin{equation}
    \mathbf{r}(t) = x(t)\mathbf{i} + y(t)\mathbf{j} + z(t)\mathbf{k} , \quad t \in [a,b]\label{eq:curve}.
\end{equation}
or denote as $\mathbf{r}(t) = (x(t),y(t),z(t))$,
 and we assume the parametric curve $\mathbf{r}(t)$ is referred to be a vector valued function with at least third order derivatives.

In differential geometry, the degree of touching is a measure used to describe the consistency of two curves at their intersection.
\begin{definition}
Assume curves $C_1$ and $C_2$ have a intersection point $p_0$.
Furthermore, 
$p_1$ and $p_2$ are points at $C_1$ and $C_2$, respectively with arc length $\Delta s$ away from $p_0$. If there exists an positive integer $n$ satisfying
\begin{equation}
    \lim_{\Delta s \to 0} \frac{|p_1 p_2|}{(\Delta s)^n} = 0 , \qquad \lim_{\Delta s \to 0} \frac{|p_1 p_2|}{(\Delta s)^{n+1}} \neq 0,
\end{equation}
then we call that the intersection point $p_0$ of $C_1$ and $C_2$ has n-degree touching.
\end{definition}
In fact, a regular curve and its n-degree Taylor expansion have at least n-degree touching according to theorem \ref{thm:talyor-touching} in Appendix.

\subsection{minimal surface}
A parametric surface $S$ can be represented a continuous map as follows
\begin{gather}
    S: D \to E^3,\quad D\subset E^2, \\
    S(u,v) = (x(u,v),y(u,v),z(u,v)), 
\end{gather}
where $(u,v) \in D$.

In general, a surface with mean curvature equal to zero is called a minimal surface. But we have an intuitive conclusion about the minimal surface.
If the area of the surface $S$ is a minimum value in the collection $M$ of surfaces with the same boundary curve $C$, this is equal to 
\begin{equation}
    S = {\arg \min}_{S' \in M}\{ area(S') | \partial S' = C \}.
\end{equation}
Then the surface $S$ must be the minimal surface.

\section{Methods}
\begin{figure*}[!htbp]
  \centering
  \includegraphics[width= 0.6\paperwidth]{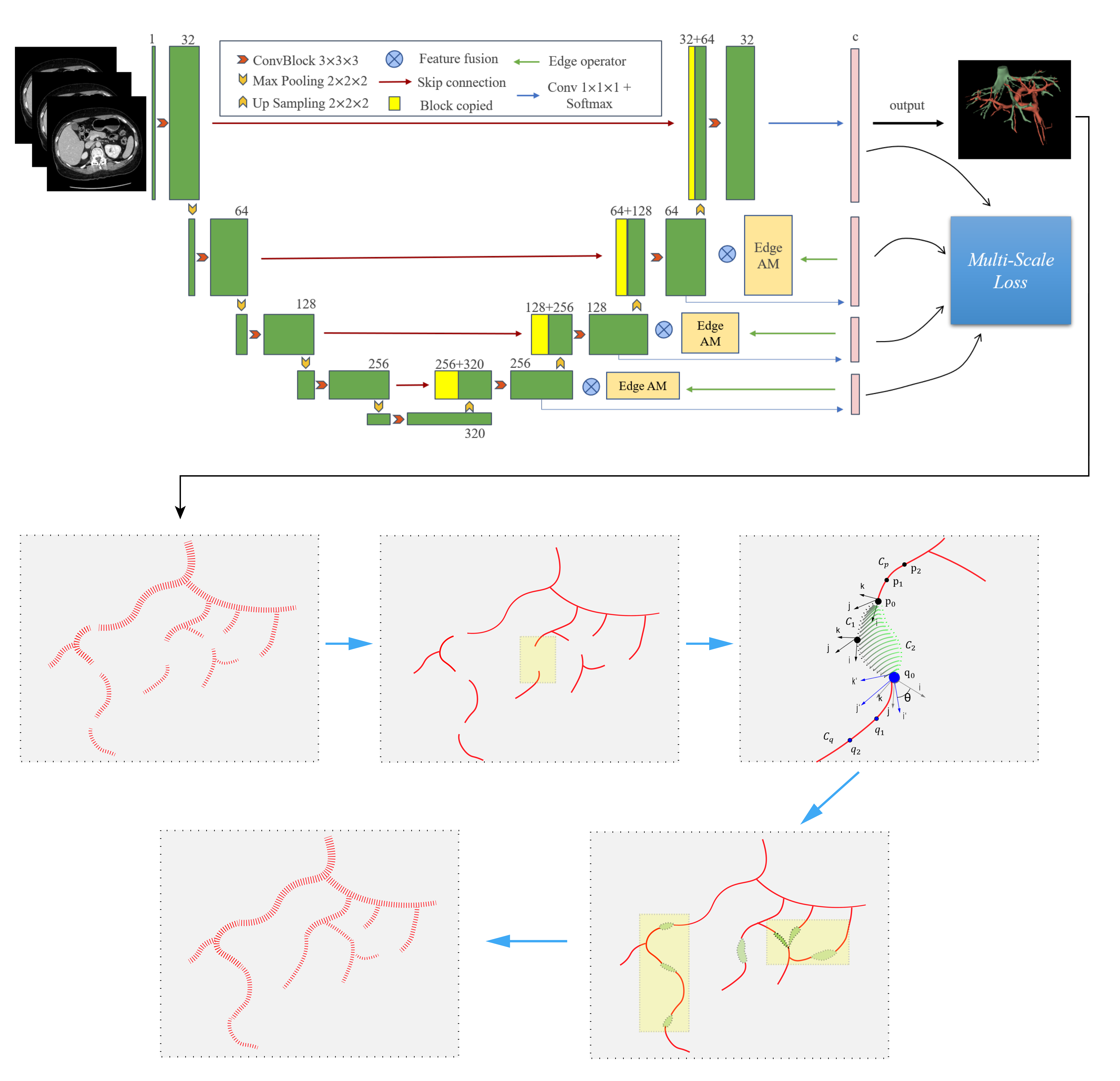}
  \caption{The overall flowchart of the proposed method, which includes two main stages. }
  
  \label{fig:overall}
\end{figure*}

Our method consists of two stages.
The first stage uses the MEA-Net for vessel segmentation, which improves the network's ability to distinguish vessels edge. 
In the second stage, The OGMC model corrects the fragments of vessel segmentation from the first stage.
The general flow of our method is shown in Figure \ref{fig:overall}. 

\subsection{Morphology edge attention network}
Because vascular boundaries often exhibit low contrast, most existing algorithms suffer much from to fragmented segmentation results. To address this issue, we propose a solution called MEA-Net, as shown in Figure \ref{fig:overall}, which incorporates an edge attention module into the U-Net \cite{UNet2015} architecture. 
This network is designed to focus on the edges of multi-scale predictions and facilitate deep supervised training. Furthermore, the attention module has a clear physical interpretation.

\subsubsection{Edge attention module}
\begin{figure*}[!htbp]
    \centering
     \subfloat[Edge attention module]{
        \includegraphics[width= 0.23\paperwidth]{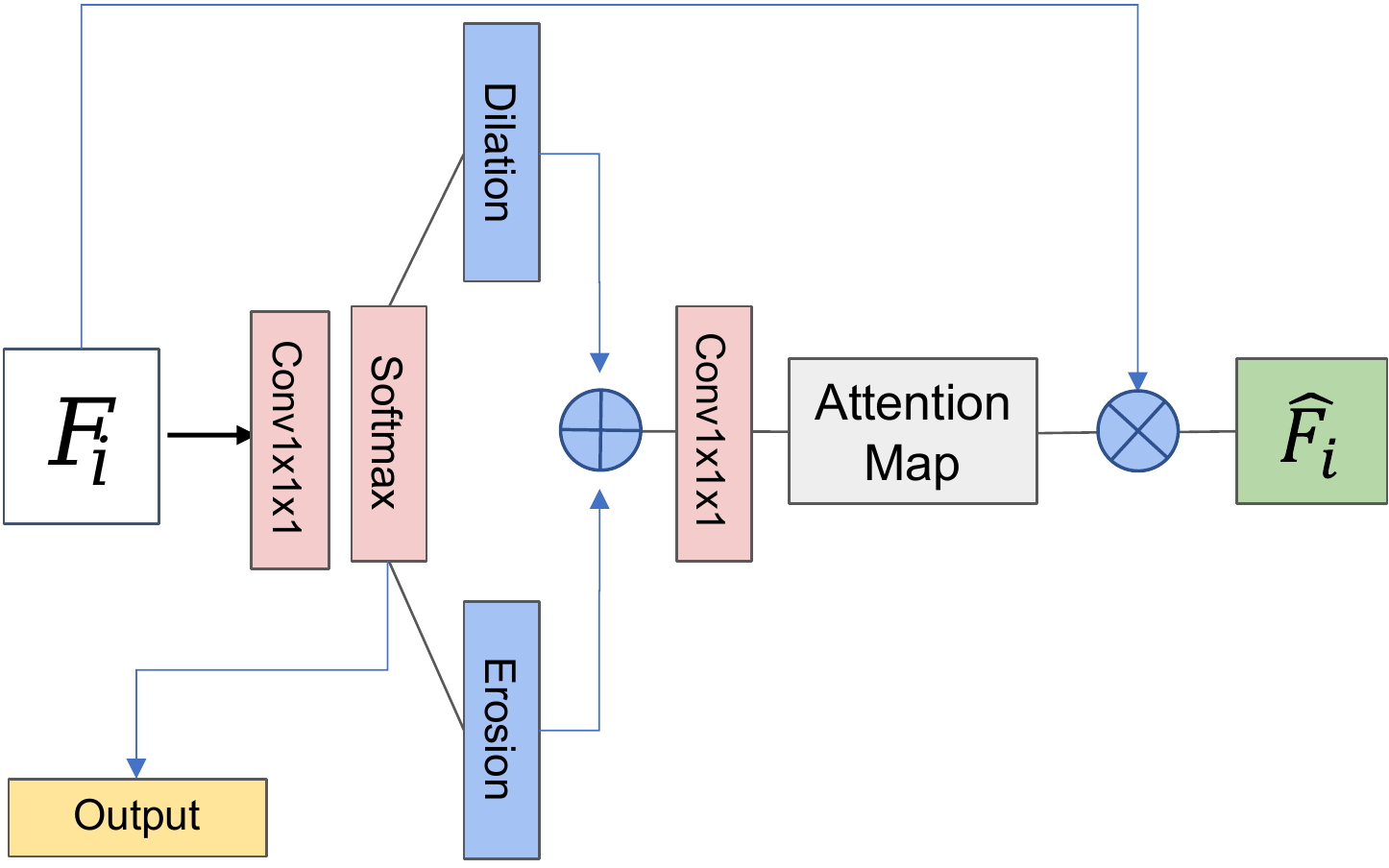}
    }
    \subfloat[Processes of visualization extracting boundary]{
        \includegraphics[width= 0.23\paperwidth]{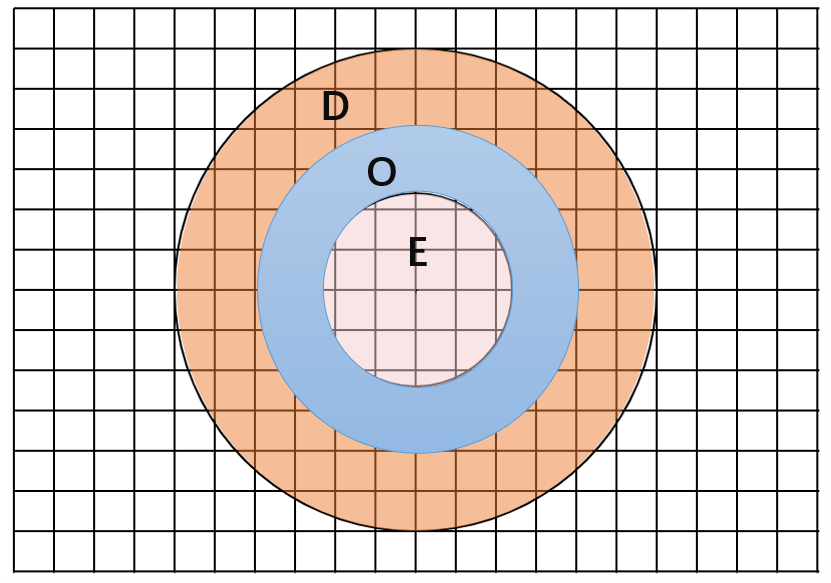}
    }
    \caption{
    (a) The forward process of extracting the boundaries with soft morphological operators for the output.
    (b) $D$ is dilation of $O \cup E$. $E$ is erosion of $O \cup E$. 
    $D \backslash E$ is edge.
    }
    \label{fig:edgeam}
\end{figure*}

The module's forward process is illustrated in Figure \ref{fig:edgeam}. The input to the module is a feature $F_i\in R^{c \times h \times w \times d}$ generated at the $i$-th decoder layer, where $h$, $w$, $d$ and $c$ represent the height, width, depth, and number of channels of the feature map, respectively.
After a $1\times 1 \times 1$ convolutional layer $Conv$, the output $S_i$ is then obtained by a $Softmax$ layer.
\begin{equation}
    S_i = Softmax( Conv(F_i))
\end{equation}

We use a differentiable erosion operator $E_{op}$ \cite{shitClDiceNovelTopologyPreserving2021} and a dilation operator $D_{op}$ to act on $S_i$ as shown in Figure \ref{fig:edgeam}. 
For simplicity, let us assume that $S_i^k$ is a binary mask,
We define $E=E_{op}(S_i^k)$ as the erosion of $S_i^k$, $O = S_i^k - E$,
and $D=D_{op}(S_i^k)$ represents the dilation of $S_i^k$.
If we view $E$ and $D$ as sets of points having value equal to 1, then $S_i^k = O \cup E$ is the foreground of the prediction. 
It can be extended to $S_i$.
So, after performing a $1\times 1 \times 1$ convolution, the attention map is obtained using
\begin{equation}
    A_i = Conv(|D_{op}(S_i) - E_{op}(S_i)|).
\end{equation}
The boundary features are then obtained by the Hadamard product as shown in the equation:
\begin{equation}
    E_i = A_i * F_i .
\end{equation}
Finally, the boundary features are fused with $F_i$ to obtain new features that will replace $F_i$ as input to the next decoder layer. This is specified by
\begin{equation}
    \hat{F_i} = F_i + E_i.
\end{equation}

\subsubsection{Mutil-scale loss}
The total loss is obtained by summing the loss functions of all the output layers of the decoder. This can be expressed as
\begin{gather}
    L_{layer} = L_{ce} + L_{dice},  \\
    L_{total} = \sum^{4}_{i=1} L_{layer}(S_i,G_i) , 
\end{gather}
where $L_{ce}$ and $L_{dice}$ denote the cross-entropy and dice loss, and $G_i$ is the corresponding ground truth with the same size as $S_i$.

\subsection{Optimal geometric matching connection model}

Segmentation of blood vessels using network methods often results in broken vessels, which alter their overall shapes and topology structures. To address this issue, we propose the OGMC model to identify ruptured vessel endpoints and optimal connection pairs to revise the segmentation produced by the first stage. The entire algorithm workflow is illustrated in Figure \ref{fig:overall}, and algorithm \ref{tab:my_algorithm}. 
There are four steps in total.

\subsubsection{Discrete tree representation of vessels }
\begin{figure}
    \centering
    \includegraphics[width= 0.6\linewidth]{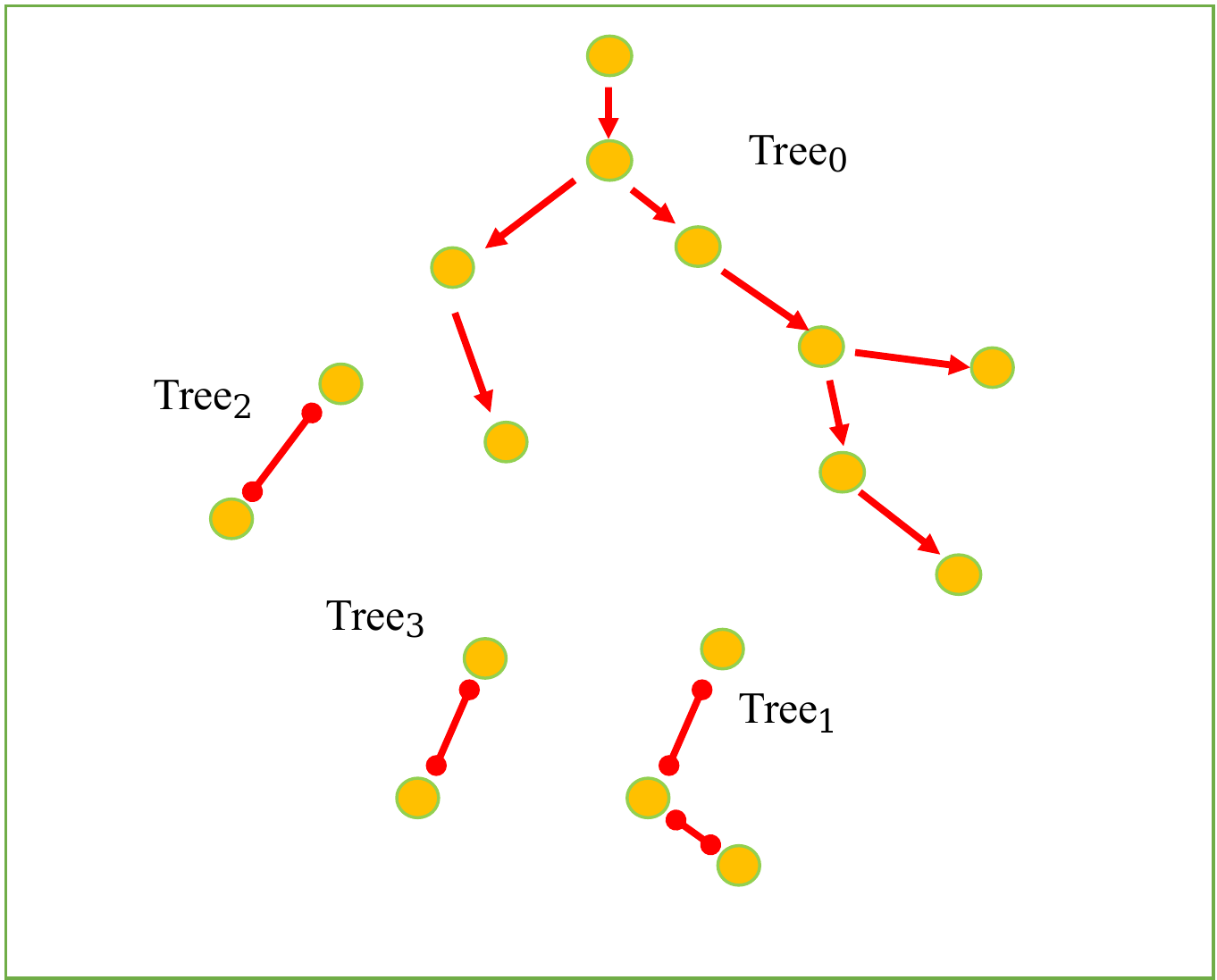} 
    \caption{Skeleton vessels and Building the vessel-tree.$Tree_1$ is main vessel-tree, we can get its direction. Other sub-tree is bi-direction.}
    \label{fig:my_skeleton}
\end{figure}
Obtaining a vessel tree involves several steps.
First, we use the algorithm \cite{LEE1994462} to thin the vessel segmentation into a set of 1-dimension skeletons $\{skel_{i}\}$, where $skel_{i}$ denotes a connected component of the skeletons. 
Second, we extract the vascular skeleton point set from the skeletons, which includes endpoints and bifurcation points. 
Next, the largest skeleton $skel_0$ is used to construct a directed vascular tree, denoted as $T_{main} = <V,E>$. 
Additionally, each other skeleton $skel_j$ creates an undirected vascular tree that is constructed as $T_{i} = <V_{i},E_{i}>$. Figure \ref{fig:my_skeleton} illustrates this process. Here, $V$ and $E$ represent the skeleton points and edges of $T_{main}$, respectively, and $V_{i}$ and $E_{i}$ represent the skeleton points and edges of $T_{i}$.

\subsubsection{Connect condition based on touching}
After we have a set of vessel trees, how can we find the pairs of endpoints needed to be connected? 
Indeed, from fluid dynamics, we can assume that if a pair of endpoints is generated by a broken vessel, their flow orientation must be consistent. This is a geometry problem. We introduce the concept of curve touching rooted in differential geometry, which measures the degree to which two intersecting curves depart from each other near their intersection point.


Now, let us concentrate on a local fragment of segmentation.
There is a pair of disjoint vessel curves $\{C_p; p_0,p_1, p_2\}$ and $\{C_q; q_0,q_1, q_2\}$, where $p_0,p_1, p_2 \in C_p$ $ q_0,q_1, q_2 \in C_q$, and $p_0,q_0$ is endpoint of $C_p,C_q$, respectively.
So we need to find a most suitable curve to connect $p_0$ and $q_0$.
Thanks to the theorem \ref{thm:curve} of the local theory of curves, see Appendix, we can fit any curve $\mathbf{r}(s)$ by a three order differentiable curve in local.

Do Taylor expansion on the regular curve $\mathbf{r}(s)$ at $s=0$,
\begin{equation}
    \mathbf{r}(s) = \mathbf{r}(0) + s\dot{\mathbf{r}}(0) + \frac{s^2}{2!}\ddot{\mathbf{r}}(0) + \frac{s^3}{3!}\dddot{\mathbf{r}}(0) + \mathbf{o}(s^3).
\end{equation}
We take the Frenet frame $\{ \mathbf{r}(0);\alpha(0), \beta(0), \gamma(0)\}$ of curve at $s=0$ to be the origin of the Cartesian coordinate system in space $E^3$,
and the first three terms of Taylor expansion as the main part.
So we can use below the parametric equation of curve to fit the main part.
\begin{align}
    \begin{cases}
    x(s) = s\\
    y(s) = \frac{\kappa_0}{2}s^2 \\
    z(s) = \frac{\kappa_0 \tau_0}{6}s^3
    \end{cases}
    \label{eq:fit}
\end{align}
Where $\kappa_0, \tau_0$ are constant parameters.
The approximation curve has the following properties.
\begin{proposition}
The equations \eqref{eq:fit} representing three-order differentiable curve $\hat{\mathbf{r}}(s) $ have the same curvature, torsion, and Frenet frame with $\mathbf{r}(s) $ at $s=0$. Thus $\hat{\mathbf{r}}(s) $
is a local approximate to the curve $\mathbf{r}(s)$.
\end{proposition}
A three-order curve $C_1$ is fitted by the equation \eqref{eq:fit} using the points 
$(p_1,p_0,q_0)$ 
to connect the curves $C_p$ and $C_q$. In this case, both the fitted curve $C_1$ and the missing curve have the same endpoints ($p_0$ and $q_0$), curvature, torsion and Frenet frame $\{ C_1; \alpha, \beta, \gamma\}$, indicating their equivalence based on theorem \ref{thm:curve}. 

To measure the consistency quantitatively, we move the frame $\{C_1; \mathbf{i}=\alpha(p_0), \mathbf{j}=\beta(p_0), \mathbf{k}=\gamma(p_0)\}$ in parallel along the curve $C_1$ to the point $q_0$. This means that the covariant differentiation of the frame with respect to the curve $C_1$ is zero:
\begin{equation}
    \nabla_{dC_1/ds}\mathbf{i} = \nabla_{dC_1/ds}\mathbf{j} =\nabla_{dC_1/ds}\mathbf{k} = 0.
\end{equation}
Next we calculate the touching bias $D_{p}(q_0)$ between the frame $\{C_1;\mathbf{i}(q_0), \mathbf{j}(q_0), \mathbf{k}(q_0)\}$ and the frame $\{ C_q; \mathbf{\hat{i}}=\hat{\alpha}(q_0), \mathbf{\hat{j}}=\hat{\beta}(q_0), \mathbf{\hat{k}}=\hat{\gamma}(q_0)\}$ of the curves $C_1$ and $C_q$ at point $q_0$.
\begin{equation}
    \begin{split}
       &\ D_{p}(q_0) = \frac{1}{3} (\|\mathbf{i}(q_0) - \mathbf{\hat{i}}(q_0) \|_{L^2} + \\
    &\ \|\mathbf{j}(q_0) - \mathbf{\hat{j}}(q_0) \|_{L^2} 
    +  \|\mathbf{k}(q_0) - \mathbf{\hat{k}}(q_0) \|_{L^2}).
    \label{eq:touching}
    \end{split}
\end{equation}

Similarly, we fit a curve $C_2$ using the points 
$(q_1,q_0,p_0)$ 
to connect $q_0$ and $p_0$. This allows us to calculate the touching bias $D_{q}(p_0)$.
Next, we introduce the touching fit degree as a quantitative measure to determine whether there is a connection relationship between $p_0$ and $q_0$.

\begin{definition}
The touching fit degree (TFD) of the curves $C_p$ and $C_q$ is as follows,
\begin{equation}
    TFD(p_0,q_0) = \frac{(D_{p}(q_0) + D_{q}(p_0))}{2}
\end{equation}

\end{definition}
Thus, by setting an upper bound $\epsilon = \sqrt{2}$, if $TFD < \epsilon $, it indicates that the two curves $C_p$ and $C_q$ are supposed to be connected.

\subsubsection{vessel global optimal connect by minimal surfaces }
From the local to the overall, 
we have a candidate set of point pairs 
$$J = \{ (p_i^j, p_l^k)| TFD(p_i^j, p_l^k) < \epsilon, p_i^j \in endpoints(T_i) \},$$ 
and a point can be in pairs with multiple points, so
what is the optimal order of connection for all point pairs within the vascular tree? 

While the touching fit degree provides valuable information about how well two curves fit based on their derivatives, it does not take into account additional factors such as the distance between nodes. 
So we propose the minimal surface matching order of the point pairs.
\begin{definition} 
Given the mutually separated curve pairs $C_p$ and $C_q$, and the fitted curve pairs $C_1$ and $C_2$. We call Msmo \eqref{eq:my_area} is the minimal surface matching order value of $C_p$ and $C_q$.
\begin{gather}
    Msmo(C_p,C_q) = \frac{1}{area(S)}, \label{eq:my_area} \\
    S = {\arg \min}_{\bar{S} }\{area(\bar{S}) |\partial \bar{S} = C_1 \cup C_2 \}.
\end{gather}
\end{definition}
For example, in Figure \ref{fig:overall}, the two fitted curves $C_1$ and $C_2$ have the same endpoints, $q_0$ and $p_0$. By combining these curves, we can form a closed curve $\Gamma = C_1 \cup C_2$. The area of the minimal surface bounded by $\Gamma$ can be used to quantify the dissimilarity between the two curves. Smaller surface areas indicate a better fit between the curves and therefore a higher reliability, which implies a higher priority in the connection process.
In situations where a point can be paired with multiple points, we can select the optimal matching pair. This ensures that the most appropriate connections are made within the vascular tree.

\subsubsection{Tracking centerlines of vessels}
Once we have identified the endpoints to connect, we can use the geodesics to be possible paths, which are minimal paths in the context of a particular Riemannian metric $\mathcal{M}$. Our purpose is to find the minimal path between two endpoints, $p^*$ and $q^*$, by minimizing the following problem:
\begin{align}
    \inf_{\mathbf{r}} \int_0^1 \sqrt{\langle \dot{\mathbf{r}}(s),\mathcal{M}(\mathbf{r}(s)) \dot{\mathbf{r}}(s) \rangle} ds,
\end{align}
subject to the boundary conditions $\mathbf{r}(0)=p^*$ and $\mathbf{r}(1)=q^*$. 

To efficiently solve this minimal path problem we can employ the fast marching method as described in the references \cite{mirebeauHamiltonianFastMarching2019, benmansourFastObjectSegmentation2009, liuTrajectoryGroupingCurvature2022}. This method provides an efficient way to compute geodesic paths.
Finally, we can fill the columnar neighbourhood with the geodesic path. The average radius at both vessel endpoints can be used to determine the columnar radius, ensuring that the neighbourhood is filled appropriately.

\begin{algorithm}
    \caption{Pseudo code for the OGMC model.}
    \label{tab:my_algorithm}
\begin{algorithmic}[1]
        \STATE \textbf{Input:} Segmentation $I$. 
        \STATE  Building vessel tree $C = \{T_i |area(T_i) > area(T_{i+1}), 0 \leqslant i < n\}$ and $T_{main} = T_0$; 
        \STATE Extracting $T_i,0 \leqslant i < n$ endpoints as $P_i =\{ p_i^j|0 < j < m_i\}$; 
        \STATE $C = C\setminus T_{main}$;
        \WHILE{$C \neq \varnothing $} 
        \STATE   $ J=\varnothing $; 
        \FOR{$T_i$ \textbf{in} $C$}    
        \STATE   Computing $TFD_{jk}^i$ and minimal surface area $A_{jk}^i$ of $(p_0^j, p_i^k ),0 < j < m_0, 0 < k < m_i$;
        \IF{ $TFD_{jk}^i < \epsilon$} \STATE  $J = J \cup (p_0^j, p_i^k )$; 
        \ENDIF
        \ENDFOR
        \IF{$J == \varnothing $} \STATE \quad break;  \ENDIF
        \STATE $(p^*,q^*) = argmin\{A | (p,q)\in J  \}$;   
        \STATE  Applying geodesic to connect $(p^*,q^*)$;  
        \STATE   $P_{main} = (P_{main} \cup P^* )\setminus (p^*,q^*)$;
        \STATE   $C = C\setminus T^*$;
        
        \ENDWHILE
        \STATE   \textbf{Output:} Segmentation $\hat{I}$.
\end{algorithmic}
\end{algorithm}

\section{Datasets and segmentation evaluation}
   Pass

\ifCLASSOPTIONcaptionsoff
  \newpage
\fi



\bibliographystyle{IEEEtran}
\bibliography{IEEEabrv,reference}
\end{document}